\documentclass[twoside,twocolumn,9pt,wasysym,amsmath,amssymb]{article}
\usepackage{extsizes}
\usepackage[super,sort&compress,comma]{natbib} 
\usepackage[version=3]{mhchem}
\usepackage[left=1.5cm, right=1.5cm, top=1.785cm, bottom=2.0cm]{geometry}
\usepackage{balance}
\usepackage{times,mathptmx}
\usepackage{sectsty}
\usepackage{graphicx} 
\usepackage{lastpage}
\usepackage[format=plain,justification=justified,singlelinecheck=false,font={stretch=1.125,small,sf},labelfont=bf,labelsep=space]{caption}
\usepackage{float}
\usepackage{fancyhdr}
\usepackage{bm,amsmath,wasysym,amssymb}
\usepackage{fnpos}
\usepackage[english]{babel}
\addto{\captionsenglish}{%
  
}
\usepackage{array}

\usepackage{droidsans}
\usepackage{charter}
\usepackage[T1]{fontenc}
\usepackage[usenames,dvipsnames]{xcolor}
\usepackage{setspace}
\usepackage[compact]{titlesec}
\usepackage{hyperref}

\definecolor{cream}{RGB}{222,217,201}

\begin{document}

\pagestyle{fancy}
\thispagestyle{plain}

\fancypagestyle{plain}{

\renewcommand{\headrulewidth}{0pt}
}

\makeFNbottom
\makeatletter
\renewcommand\LARGE{\@setfontsize\LARGE{15pt}{17}}
\renewcommand\Large{\@setfontsize\Large{12pt}{14}}
\renewcommand\large{\@setfontsize\large{10pt}{12}}
\renewcommand\footnotesize{\@setfontsize\footnotesize{7pt}{10}}
\makeatother

\renewcommand{\thefootnote}{\fnsymbol{footnote}}
\renewcommand\footnoterule{\vspace*{1pt}%
\color{cream}\hrule width 3.5in height 0.4pt \color{black}\vspace*{5pt}} 
\setcounter{secnumdepth}{5}

\makeatletter 
\renewcommand\@biblabel[1]{#1}            
\renewcommand\@makefntext[1]%
{\noindent\makebox[0pt][r]{\@thefnmark\,}#1}
\makeatother 
\renewcommand{\figurename}{\small{Fig.}~}
\sectionfont{\sffamily\Large}
\subsectionfont{\normalsize}
\subsubsectionfont{\bf}
\setstretch{1.125} 
\setlength{\skip\footins}{0.8cm}
\setlength{\footnotesep}{0.25cm}
\setlength{\jot}{10pt}
\titlespacing*{\section}{0pt}{4pt}{4pt}
\titlespacing*{\subsection}{0pt}{15pt}{1pt}

\fancyfoot{}
\fancyfoot[LO,RE]{\vspace{-7.1pt}\includegraphics[height=9pt]{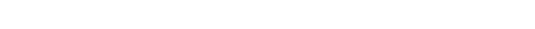}}
\fancyfoot[CO]{\vspace{-7.1pt}\hspace{13.2cm}\includegraphics{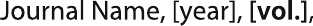}}
\fancyfoot[CE]{\vspace{-7.2pt}\hspace{-14.2cm}\includegraphics{head_foot/RF}}
\fancyfoot[RO]{\footnotesize{\sffamily{1--\pageref{LastPage} ~\textbar  \hspace{2pt}\thepage}}}
\fancyfoot[LE]{\footnotesize{\sffamily{\thepage~\textbar\hspace{3.45cm} 1--\pageref{LastPage}}}}
\fancyhead{}
\renewcommand{\headrulewidth}{0pt} 
\renewcommand{\footrulewidth}{0pt}
\setlength{\arrayrulewidth}{1pt}
\setlength{\columnsep}{6.5mm}
\setlength\bibsep{1pt}

\makeatletter 
\newlength{\figrulesep} 
\setlength{\figrulesep}{0.5\textfloatsep} 

\newcommand{\topfigrule}{\vspace*{-1pt}%
\noindent{\color{cream}\rule[-\figrulesep]{\columnwidth}{1.5pt}} }

\newcommand{\botfigrule}{\vspace*{-2pt}%
\noindent{\color{cream}\rule[\figrulesep]{\columnwidth}{1.5pt}} }

\newcommand{\dblfigrule}{\vspace*{-1pt}%
\noindent{\color{cream}\rule[-\figrulesep]{\textwidth}{1.5pt}} }

\makeatother

\twocolumn[
\begin{@twocolumnfalse}
\begin{center}

\noindent\LARGE{Oscillations of a cantilevered micro beam driven by a viscoelastic flow instability} \\ 
\vspace{0.6cm}
\noindent\large{Anita A. Dey,\textit{$^{a}$}\textit{$^{b}$} Yahya Modarres-Sadeghi,\textit{$^{a}$} Anke Lindner,\textit{$^{b}$} and Jonathan P. Rothstein\textit{$^{a}$}} \\
\end{center} \vspace{0.5cm}
\noindent\normalsize{The interaction of flexible structures with viscoelastic flows can result in very rich dynamics. In this paper, we present the results of the interactions between the flow of a viscoelastic polymer solution and a cantilevered beam in a confined microfluidic geometry. Cantilevered beams with varying length and flexibility were studied. With increasing flow rate and Weissenberg number, the flow transitioned from a fore-aft symmetric flow to a stable detached vortex upstream of the beam, to a time-dependent unstable vortex shedding. The shedding of the unstable vortex upstream of the beam imposed a time-dependent drag force on the cantilevered beam resulting in flow-induced beam oscillations. The oscillations of the flexible beam were classified into two distinct regimes: a regime with a clear single vortex shedding from upstream of the beam resulting in a sinusoidal beam oscillation pattern with the frequency of oscillation increasing monotonically with Weissenberg number, and a regime at high Weissenberg numbers characterized by 3D chaotic flow instabilities where the frequency of oscillations plateaued. The critical onset of the flow transitions, the mechanism of vortex shedding and the dynamics of the cantilevered beam response are presented in detail here as a function of beam flexibility and flow viscoelasticity.} \\
\end{@twocolumnfalse} \vspace{1cm}

]


\renewcommand*\rmdefault{bch}\normalfont\upshape
\rmfamily
\section*{}
\vspace{-1cm}


\footnotetext{\textit{$^{a}$~Department of Mechanical and Industrial Engineering, University of Massachusetts, Amherst, Massachusetts 01003, USA. E-mail: rothstein@ecs.umass.edu}}
\footnotetext{\textit{$^{b}$~Laboratoire de Physique et M\'{e}canique des Milieux H\'{e}t\'{e}rog\'{e}nes, UMR 7636, CNRS, ESPCI Paris, PSL Research University, Universit\'{e} Paris Diderot, Sorbonne Universit\'{e}, Paris 75005, France.}}



\section{Introduction}

Fluid-structure interactions (FSI) have been heavily studied by researchers because of their ubiquity in a variety of mechanical, industrial and biological processes. At high Reynolds numbers, the interaction of flexible structures with flow instabilities leads to very rich dynamics documented in many books and review papers \cite{bearman1984,blevins1990flow,paidoussis1998fluid,paidoussis2004fluidv2, paidoussis2011crossflow, Sarpkaya2004389,williamson2004vortex}. In low Reynolds number flows, although flows are stable, complexity arises from non-linear interactions between deformable structures and viscous flow. Viscous fluid motion can modify the shape, orientation and position of a structure which in turn leads to coupling between the flow field and the structural response \cite{du2019dynamics}. FSI studies of these flows are relevant to the biological and physiological world seen in the flow past flagella \cite{brennen1977fluid}, swimming of micro-organisms \cite{Lauga_2009}, the deformation of red blood cells during transport in blood vessels \cite{Abkarian2016} or the deformation of soft fluid-conveying vessels \cite{heil2011fluid,bertram2003experimental}. An important class of low Reynolds number flows includes viscoelastic fluid flows. In these flows, purely elastic instabilities can occur even in the absence of inertia 
\cite{shaqfeh1996,mckinley1996rheological,larson1992instabilities,groisman2000elastic} and can in turn interact with flexible structures. Although these elastic flow instabilities have been reported in a host of viscoelastic fluids and flow geometries, such viscoelastic FSI studies (VFSI) remain scarce and have only recently been conducted for the flow of wormlike micelle solutions past flexible structures placed in a crossflow \cite{dey2017,dey2018}. However, the interplay between the various types of viscoelastic fluids, flexible structures and flow geometries is expected to lead to a large variety of dynamics, relevant for a number of fields such as low Reynolds number flows and structural mechanics.    

Polymer solutions are often classified as viscoelastic fluids due to the complex behavior of these fluids imparted by the physical nature of a mobile polymer macromolecule. As a flexible polymer coil stretches within a flow field, it is deformed out of its equilibrium random walk configuration. An elastic restoring force results, driving the polymer back toward its entropically favorable equilibrium state \cite{Flory1953}. High molecular weight polymers can thus impart an entropic elasticity to a fluid which allows the polymer solution or melt to carry stress along the flow streamlines and can lead to the build up of normal stresses in simple shear flows. The importance of elasticity in the flow is described by the non-dimensional Weissenberg number, $ Wi $ = $ \lambda\dot{\gamma} $, where $\lambda$ is fluid relaxation time and $\dot{\gamma} = U/L$ is the shear rate,  where $U$ is the flow velocity and $L$ is the characteristic lengthscale. The importance of inertia is described by the Reynolds number, $Re = UL/ \nu $, where $\nu$ is the kinematic viscosity of the fluid. High Weissenberg number flows have become easily achievable in the absence of inertia, $Re \ll 1$, in microfluidics \cite{microfluidics}. The micrometer sized flow geometries result in large shear rates \cite{RODD20051,pipe2009microfluidic} and Weissenberg numbers while simultaneously minimizing the Reynolds number. The combination of large elastic stresses resulting from those high Weissenberg numbers and streamline curvature, \cite{mckinley1996rheological,pathak2004elastic} while keeping the Reynolds number small, leads to purely elastic flow instabilities, making these flows quite different from Newtonian fluid flows. The low Reynolds number-high Weissenberg number space that can be probed due to microfluidics has led to many studies of viscoelastic instabilities in polymer solutions. Various microfluidic geometries such as contraction-expansion flows \cite{rodd2007role}, cross-slot flows \cite{Haward2016, xi2009mechanism}, T-channel flows \cite{soulages2009investigating}, flow past cylinders \cite{Haward2019} and serpentine channel flows \cite{zilz2012} have been studied. Although all of these flows demonstrate the onset of  elastic instabilities at large Weissenberg numbers, in all of the examples, the structure geometry is rigid and not actively interacting with the instabilities. 
\begin{figure*}
\centering
\includegraphics[width=0.65\textwidth,keepaspectratio]{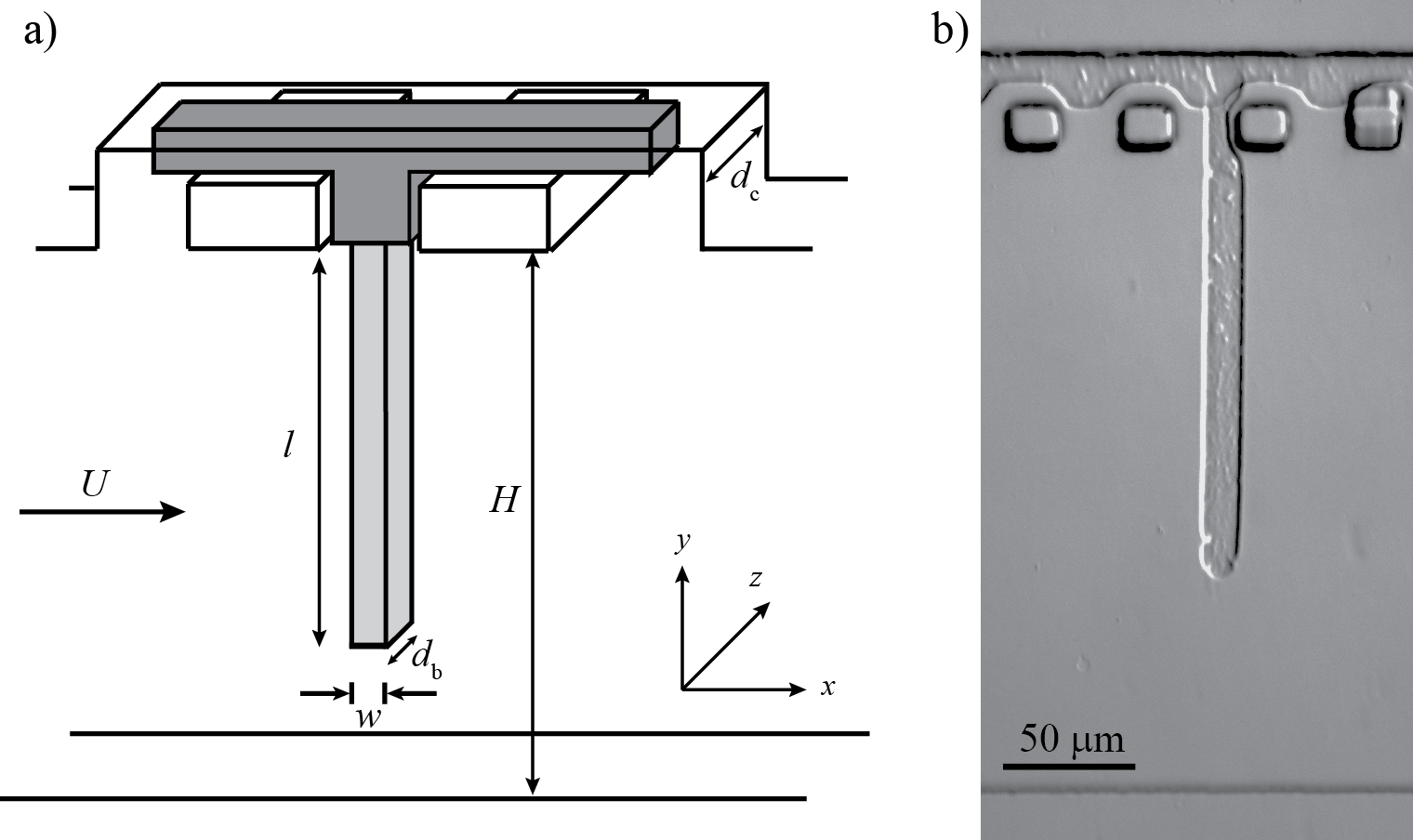}
\caption{(a) Schematic of the cantilevered beam in the fabricated microchannel and (b) an image of the fabricated beam.}
\label{setup}
\end{figure*}

\begin{table*}
\caption{\label{table} Fabricated beam geometries. The channel depth, cantilevered beam width and depth, and elastic modulus were $d_{\text{c}}=48 \, \pm \, 2 \; \mu $m, $w = 17 \, \pm \, 2 \; \mu $m and $d_{\text{b}}=40 \, \pm \, 2 \; \mu $m and $E = 3 \, \pm \, 0.2$ MPa respectively while the fluid viscosity was $\eta_{0} = 22$ mPas.}
\centering
\begin{tabular}{lcccccc} 
\hline
& \textit{l} [$\mu$m] & \textit{H} [$\mu$m] & $\alpha$ & $\kappa$ [N/m] & $\tilde{\mu}$ \\ \hline
Beam 1 & $105 \pm 2$ & $150 \pm 2$ & 0.7 & $2 \times 10^{-1}$ & $2 \times 10^{-4}$ to $1 \times 10^{-2}$ \\
Beam 2 & $290 \pm 2$ & $360 \pm 2$ & 0.8 & $1 \times 10^{-2}$ & $1.2 \times 10^{-3}$ to $1 \times 10^{-1}$  \\ \hline
\end{tabular}
\end{table*}

Alternatively, allowing these instabilities to affect the position, deformation and motion of an object falls in the domain of fluid-structure interaction problems and introduces a host of interesting questions regarding the dynamics and elasticity of the structures and boundaries of the flow. In this paper, using viscoelastic flow past a flexible beam attached as a cantilever to one side of a microchannel, we provide evidence that elastic flow instabilities occurring at high Weissenberg numbers in a confined flow can, given the right structural flexibility, indeed generate motion in the structure and subsequently couple with the structural motion. 

The geometry of thin flexible fibers used in this study has applications in the field of microfluidics in the development of micro- and nano-devices such as flow rate sensors and actuators \cite{attia2009soft,cheri2014real,li2012superelastic, Pham2015}. Development of fluid-actuated cantilevered microscale beams to act as fluid energy harvestors is also a promising technology \cite{elahi2018review}. The results of this work could provide an insight into the use of viscoelastic flow instability as a mechanism of inducing vibrations in micro-structures. Thin flexible fibers are also commonly seen in biology, where they appear as cilia and flagella used for locomotion and feeding of various micro-organisms. Micro-organisms interact with biological flows which are often viscoelastic and always at low Reynolds numbers \cite{Lauga_2009,Fu2009}. Mimicing of micro-organisms using bio-inspired cilia and flagella has applications in the development of artifical micro-swimmers, micro-pumps, valves and mixers \cite{kongthon2011dynamics,sareh2013swimming, Alvarado2017}. Our study of thin cantilevered micro-scale beams placed in viscoelastic flow could help in understanding the dynamics and interactions of micro-organisms in a host of different micro-environments.  

This paper describes the investigation of the flow of a viscoelastic polymer solution past a cantilevered beam attached to a side wall of a micro scale flow channel. The beam is confined by the top and bottom walls of the channel, occupying nearly the full channel depth and partially blocks the channel height due to its significant length. With increasing flow velocity, elastic instabilities arise in the flow due to the presence of the cantilevered beam which in turn begin to couple with the cantilevered beam. This interaction has been studied by varying the flexibility of the cantilevered beam placed in the flow path. The critical onset of beam oscillations, underlying mechanism of the oscillations, the characteristics of the flow instabilities and flow-induced beam deformation over a range of Weissenberg numbers are discussed. Across the tests of the cantilevered beams, two distinct regimes of the oscillatory response of the flexible beam are identified in this paper. 
\begin{figure*}
\centering
\includegraphics[width=0.85\textwidth,keepaspectratio]{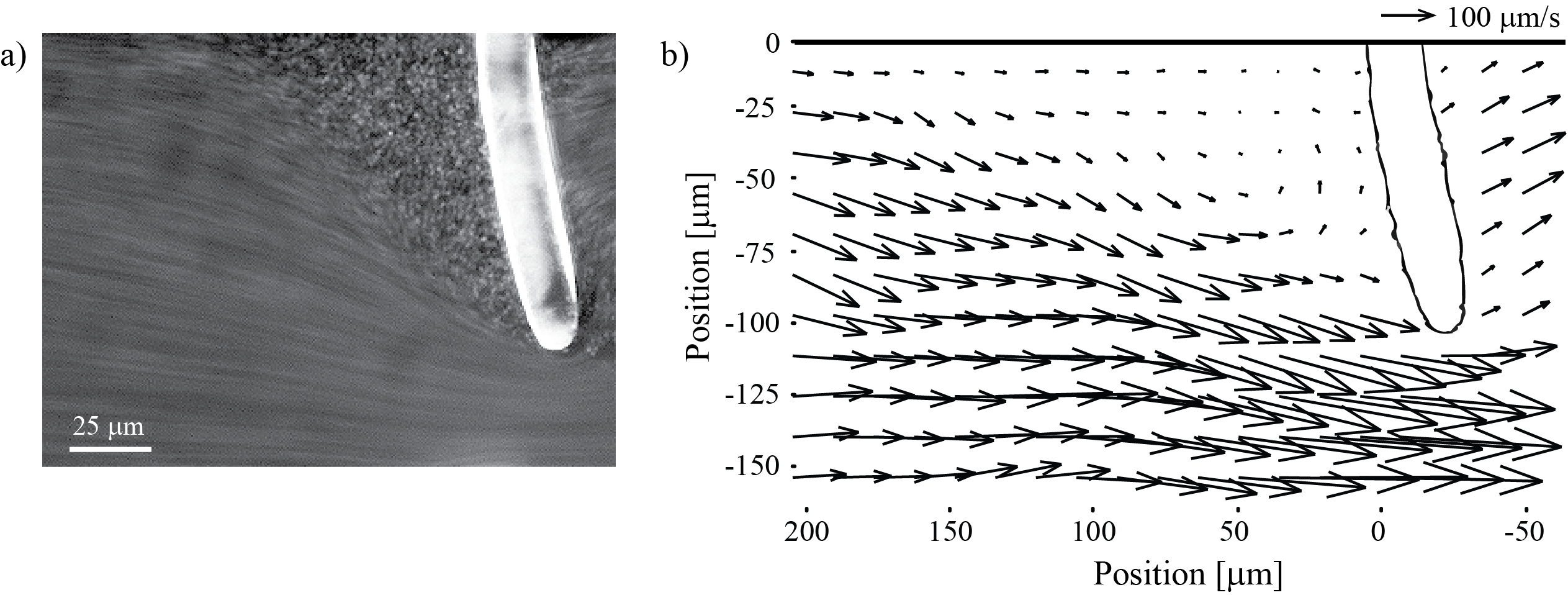}
\caption{The (a) darkfield streakline and (b) PIV images of viscoelastic flow past Beam 1 for Weissenberg numbers of (a) $Wi = 1.5$ and (b) $Wi = 3$. The flow is from left to right.}
\label{rigid}
\end{figure*}
\section{Experimental Setup}
The experimental geometry consisted of a long rectangular channel with a rectangular beam extending from one of the sidewalls (Fig. \ref{setup}(a)). The channel was made of polydimethylsiloxane (polydimethylsiloxane, Sylgard 184, Corning) (PDMS) and fabricated using traditional soft-lithography techniques with a depth of $d_{\text{c}}=48 \; \mu $m and heights of $H = 150 \; \mu $m and $360 \; \mu $m. Beams of a controlled geometry were fabricated using the stop-flow microscope-based
projection photo-lithography process \cite{Dendukuri, Berthet2016}. This method was used to fabricate beams of varying elastic moduli and flexibility \cite{Capello,Duprat2014}. The microchannel was filled with a photosensitive solution composed of $10\%$ of Darocur 1173 photo-initiator (PI, 2-hydroxy-2-methylpropiophenone, Sigma), polyethylene glycol-diacrylate (PEGDA, Mw = 700, Sigma) and a solvent (mixture of water and polyethylene glycol (PEG,Mw = 1000, (Sigma) at a ratio of 1:2 in volume) at varying proportions. Under a zero flow condition, a photomask with the beam geometry was placed in the field-stop position of a microscope (Zeiss) equipped with a UV light source (Lamp HBO 130W) and a $\times 10$ Fluar objective, and the shutter was opened for 800 ms. With an open shutter, the photosensitive solution in the microchannel exposed to UV through the photo-mask underwent polymerization. As PDMS is permeable to oxygen which quenches this reaction, a non-polymerized layer of constant thickness of approximately $5-6\; \mu$m is left along the top and bottom walls. This resulted in free-standing beams in the channel. In order to attach the beam to a wall, a polymerized wall was first fabricated around PDMS posts close to the channel walls as followed by Wexler et. al \cite{wexler2013bending}. A cantilevered beam with a clamped boundary condition was produced as shown in Fig. \ref{setup}(b).  

Cantilevered beams of varying flexibility were fabricated for the experiments conducted in this study. A rigid bounding wall was obtained by using a photosensitive solution comprised of photo-initiator and PEGDA with no additional solvent added to ensure that it remained stiff and fixed in place during the experiments. For the flexible beams, a photosensitive solution of photo-initiator and a $50\%$ aqueous solution of PEGDA was prepared. \cite{Capello}. The photosensitive solution was then flushed to introduce the viscoelastic fluid into the microchannel. As the polymerized PEGDA beam was exposed to the incoming aqueous viscoelastic solution, a swelling of the beam by $15\%$ was observed for both the relatively-rigid and flexible beams. The swelling of the beam in the microchannel was found to occur only at the initial inflow of the solution and no consequent changes in the polymerized beam were observed with time once the channel was filled with the aqueous polymer solution. Following the swelling, the elastic modulii of the $100 \%$ and $50 \%$ PEGDA solutions were measured to be 12 MPa and 3 MPa, respectively. These measurements of the elastic modulii were conducted by measuring the deflection of the fabricated beam under gravity outside of the microfluidic device. The final fabricated beams had estimated dimensions of width $w = 17 \; \mu  $m and depth $d_{\text{b}}=40 \; \mu $m and lengths of $l = 105 \; \mu$m and $l = 290 \; \mu$m depending on the experiment. The non-polymerized layer at the top and bottom of the beam was deduced to be $4\pm 1.5 \; \mu$m. The relative intensity of viscous and elastic forces acting on the beam can be compared using a  dimensionless elasto-viscous number, $\tilde{\mu} = \eta U l^{3}/ E I $, where $\eta$ is the fluid viscosity, $U$ is a typical flow velocity, $E$ is the material Young's modulus, $I$ is the area moment of inertia \cite{Capello,quennouz2015transport,du2019dynamics}. The details of the fabricated beam geometries such as  beam length, beam stiffness $\kappa = EI / l^{3}$, channel blockage ratio at zero flow conditions ($\alpha = l/H$) and the range of the elasto-viscous numbers tested are provided in Table \ref{table}. The theoretical natural frequency of the cantilevered beam in air was calculated to be $f_{N} \approx 1 $ MHz which is much larger than any oscillation frequencies that are expected in this experiment \cite{weaver1990vibration}. As a result, lock-in behavior often observed in Newtonian fluid-structure interactions is not anticipated. 

\begin{figure*}
\centering
\includegraphics[width=0.9\textwidth,keepaspectratio]{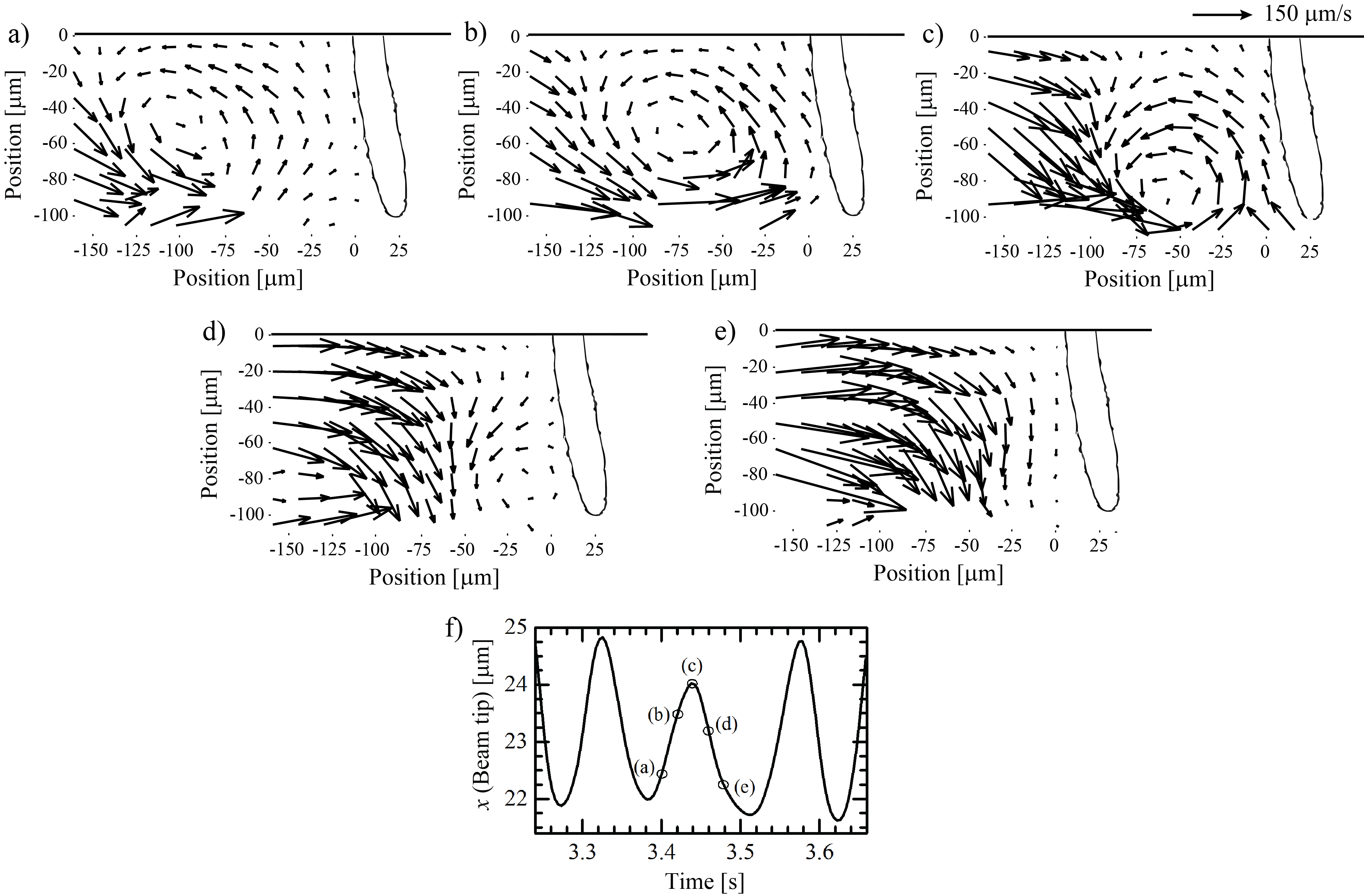}
\caption{(a-e) PIV image sequence for flow past Beam 1 at $Wi = 16$. The time interval between each two consecutive images is $20$ ms. The flow is from left to right. (f) The corresponding time history of the beam oscillations.}
\label{piv}
\end{figure*}

The test fluid was composed of Flopaam 3630 (SNF Floerger) mixed with deionized water at a concentration of 0.02 wt$\%$. Flopaam is a proprietary mixture of high molecular weight co-polymers of polyacrylamide and polyacrylate. At a concentration of 0.02 wt$\%$, the mixture showed a zero shear rate viscosity of approximately $\eta_{0} = 22$ mPas. The relaxation time of the fluid was found to be $\lambda = 0.05$ s using capillary breakup extensional rheometry experiments \cite{surcaber}. A frame rate of 500 frames per second (Hamamatsu Orca-flash 4.0 camera) was used to capture the structural motion. Particle image velocimetry was used to generate a complete and quantitative measurement of the velocity flow field around the cantilevered beam. The polymer solution was seeded with flourescent microparticles of size $1\; \mu$m (Sigma Aldrich) at $0.005\%$ by weight. A precision pump (Nemesys, Cetoni) was used to drive the flow in the microchannel at flow rates ranging from 0 to 150 nl/s. The responses of the fabricated beams and the test fluid were captured using a particle tracking software (Tracker), ImageJ (NIH) and the particle image velocimetry (PIV) technique (Lavision) respectively.   

\section{Results}
In order to illustrate the response of the beam to the oncoming flow, a dimensionless Weissenberg number, $Wi = \lambda U_{\text{gap}}/ d_{\text{c}}$, is used where $\lambda$ is the fluid relaxation time, $d_{\text{c}}$ is the channel depth and $U_{\text{gap}}$ is the average flow velocity in the gap between the tip of the beam and the opposite channel wall, neglecting leakage through the layer on top and bottom of the beam. This flow velocity in the gap is obtained as $U_{\text{gap}} = Q/(d_{\text{c}}(H - y))$, where $Q$ is the flow rate, $d_{\text{c}}$ is the channel depth and $y$ is the projected length of the deformed beam that is perpendicular to the oncoming flow. Using the gap velocity to define the Weissenberg number ensures that the maximum Weissenberg number is used and incorporates the effects of the beam deformation and a flow-rate-dependent blockage ratio ($\alpha_{\text{flow}} = y/H$) with increasing flow velocity. 

The images in Fig.~\ref{rigid} represent the streakline and PIV images captured at Weissenberg numbers of $Wi = 1.5$ and $Wi = 3$ for stable flow past Beam 1 (see details in Table \ref{table}). At these flow rates, the cantilevered beam was observed to undergo a small static deformation in the flow direction, as observed by Wexler et. al \cite{wexler2013bending} for a cantilevered beam in the flow of a Newtonian fluid. In Fig.~\ref{rigid}(a) and (b), a small re-circulation zone can be observed just upstream of the cantilevered beam. The stability of the re-circulation zone can be confirmed by the streakline image taken over the course of a long exposure time of 0.5 s. This flow separation was found to be initiated at Weissenberg numbers of $Wi \geq 1$. At these Weissenberg numbers, separated vortices upstream of flow obstacles have been observed in a number of viscoelastic microfluidic flows including flow into corners, into contractions and past posts \cite{rodd2007role,RODD20101189, hwang2017flow, shi2015mechanisms,hawardreview}. All of these flows have the combination of streamline curvature and elasticity known to be necessary for elastic vortex formation \cite{mckinley1996rheological,pathak2004elastic}. This re-circulation zone was observed to increase in size with increasing flow velocity. As seen in Fig.~\ref{rigid}, the vortex appears to originate at the corner between the beam and the upper wall. With increasing flow velocity, the vortex grows in size and intensity. Although the majority of the flow is deflected downward and around the tip of the cantilevered beam, a small fraction ($ \approx 2 \%$) of the flow can also be seen in the movies to pass through the small $4 \mu $m gap between the beam and the upper and lower walls of the microchannel. At even higher flow velocities, the vortex upstream of the beam was found to become unstable and time dependent, which in turn triggered oscillations of the beam.  
\begin{figure*}
\centering
\includegraphics[width=0.8\textwidth,keepaspectratio]{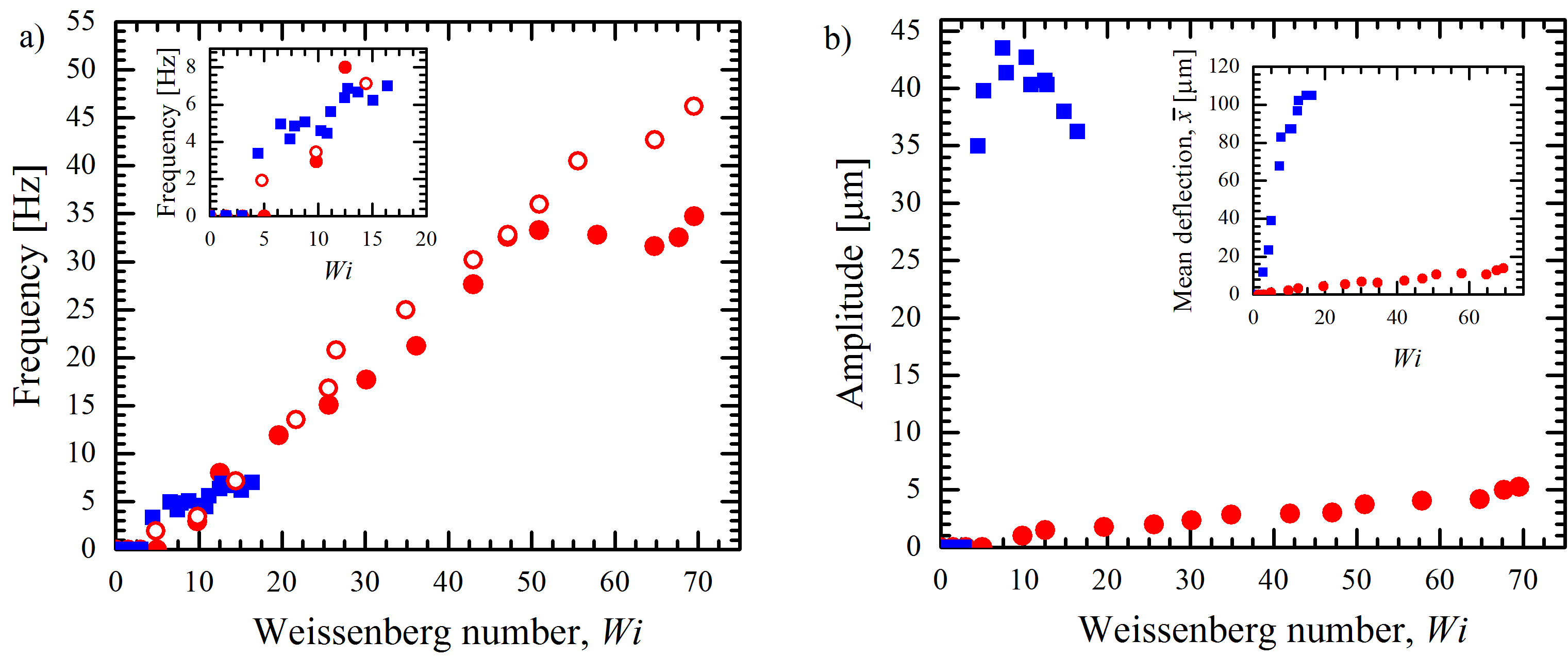}
\caption{ (a and inset) Frequency of Beam 1 ($\CIRCLE$) and Beam 2 ($\blacksquare$) along with the frequency of fluctuating velocity vectors obtained from PIV images at a point $2w = 34 \; \mu $m upstream of the tip of Beam 1 ($\Circle$) versus Weissenberg number. (b) Amplitude of Beam 1 ($\CIRCLE$) and Beam 2 ($\blacksquare$) tip oscillations versus Weissenberg number. The inset is a plot of the mean deflection ($\bar{x}$) observed during oscillations versus Weissenberg number.}
\label{plot}
\end{figure*}
This flow transition was observed to occur at a Weissenberg number of $Wi_{crit} = 5$. The shedding of the unstable vortex was observed to produce periodic beam oscillations. A sequence of PIV images captured at time intervals of $\Delta t = 20$ ms at $Wi = 16$ for a shedding vortex are presented in  Fig.~\ref{piv}. These PIV images illustrate the vortex evolution and the subsequent beam oscillations. In Fig.~\ref{piv}(a), the corner vortex at its maximum size can be seen at a location of about $100 \; \mu$m upstream of the flexible beam. As time progressed from Fig.~\ref{piv}(b) to (c), the strength of the vortex, its vorticity, is observed to increase as the vortex center began to approach the flexible beam and move from the wall. As time progresses further in Fig.~\ref{piv}(d), the center of the vortex is observed to move towards the tip of the flexible beam. At this position, the high flow velocity of the fluid passing around the tip of the cantilevered beam provides sufficient shear stress to dislodge the vortex, strip it from the beam and convect it downstream as seen in Fig.~\ref{piv}(e). In Fig.~\ref{piv}(f), the \textit{x}-position of the cantilevered beam's tip is shown as a function of time. It is clear from Fig.~\ref{piv}(f) that the growth and decay of the vortex is directly coupled to oscillations observed at the tip of the cantilevered beam. The maximum deflection of the beam tip correlates with the instance shown in Fig.~\ref{piv}(c), when the center of the vortex is at the same height as the tip of the beam, where the large torque arm helps maximize the deflection of the tip of the beam. The motion of the beam's tip in Fig.~\ref{piv}(f) appears to follow a sinusoidal motion with time. This motion is similar to the oscillations observed during vortex-induced vibration at high Reynolds numbers, but very different from the viscoelastic fluid-structure interactions previously observed for the flow of wormlike micelles past cylinders \cite{dey2018} or sheets \cite{dey2017}. In those systems, vortices upstream of the cylinder or sheet were not observed and instead, the oscillations were induced by a breakdown of the elastic fluid in the extensional flow region downstream of the cylinder or sheet. The resulting oscillations followed a saw-tooth profile as the failure of the wormlike micelles that induced the observed flow instability was quite rapid \cite{dey2018}. 
\begin{figure*}
\centering
\includegraphics[width=0.8\textwidth,keepaspectratio]{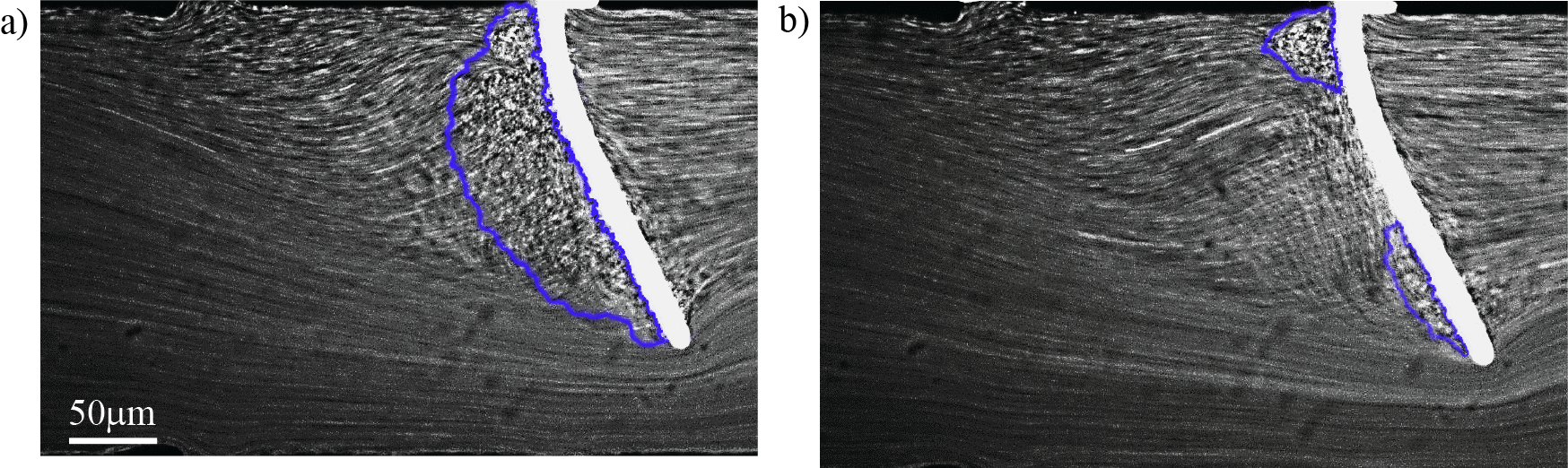}
\caption{Dark field streakline image of flow fluctuations occurring in the upstream region of Beam 2 at $Wi = 8$ for a time interval of $2.5$ ms. The flow is from left to right.}
\label{beam2}
\end{figure*}
\begin{figure*}
\centering
\includegraphics[width=0.8\textwidth,keepaspectratio]{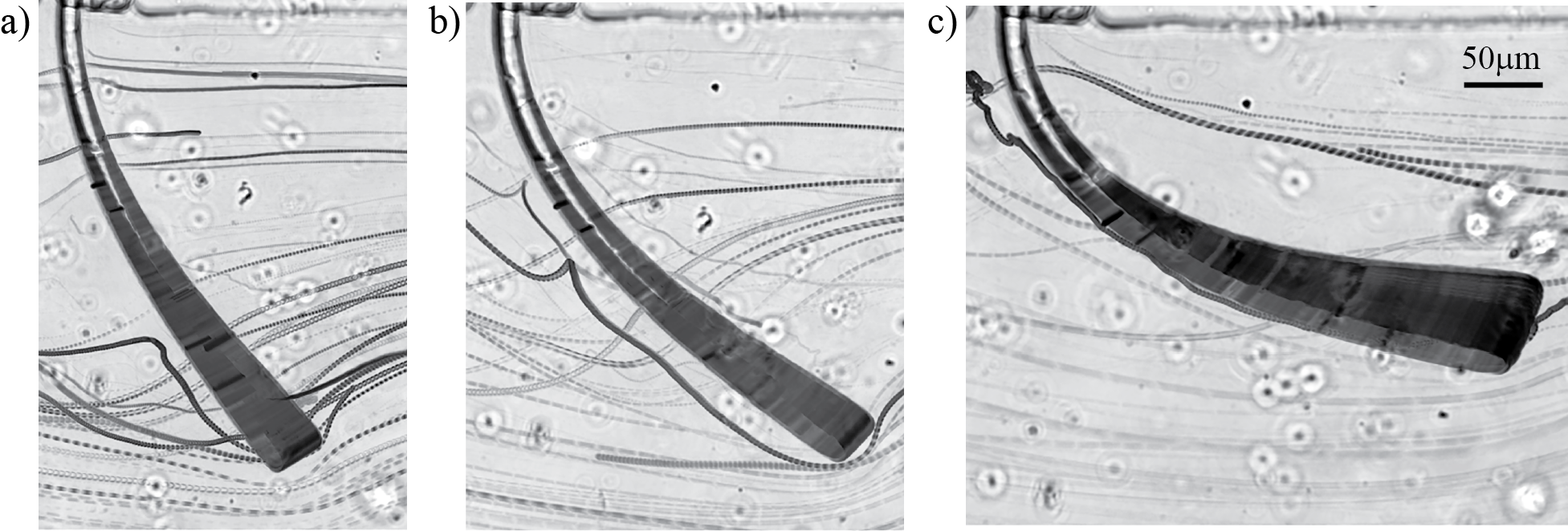}
\caption{Bright field streakline image of viscoelastic flow past Beam 2 at varying Weissenberg numbers. (a) $Wi = 5$, (b)$Wi = 8$ and (c) $Wi = 12$. The flow is from left to right.}
\label{longbeam}
\end{figure*}
In Fig.~\ref{plot}(a), the oscillation frequency of the tip of Beam 1 and the frequency of velocity fluctuations a distance of $2w = 34 \; \mu$m upstream of the tip of the beam are plotted against increasing Weissenberg number in the gap, $Wi$. The velocity fluctuations were obtained from PIV images. At the Weissenberg number of $Wi = 5$, the stable vortex upstream of the beam transitioned to an unstable time-dependent vortex resulting in a periodic shedding with a frequency of 2 Hz. The velocity fluctuations upstream of the beam resulting from the vortex shedding did not yet provide sufficient forcing to cause a time-dependent beam displacement that could be resolved at the magnification used in these experiments. A slight increase in the Weissenberg number led to the enhancement of the unstable vortex upstream of the beam and its subsequent shedding resulted in the onset of Beam 1 oscillations at the Weissenberg number of $Wi = 10$. Over most of the range where oscillations were observed, the frequency of the fluctuating velocity upstream of Beam 1 and the frequency of Beam 1 oscillations closely matched each other and increased monotonically with Weissenberg number (Fig.~\ref{plot}(a)). However, the beam oscillation frequency reached a plateau at $Wi = 50$ at about $33$ Hz while the frequency of fluctuating velocity vectors continued to increase with Weissenberg number. As the natural frequency of the beam is many orders of magnitude larger than the frequency of observed beam oscillations, $f_{\text{N}} \approx 1 $ MHz, the observed plateau is not associated with the lock-in observed for Newtonian FSI. At $Wi > 50$, the coherent vortices observed in Fig.~\ref{piv} are no longer present upstream of the beam. The oscillations beyond this Weissenberg number are induced not by the shedding of a vortex with a single dominant frequency, but by 3D chaotic-like velocity fluctuations originating upstream and observed in the flow around the tip of the beam with a dominant frequency accompanied by higher harmonics exciting the motion of the beam. This chaotic-like fluid flow around Beam 1 is akin to the flow fluctuations observed for elastic turbulence at low Reynolds numbers \cite{steinberg2000, groisman2004elastic,sousa2018purely}. This type of fluctuating flow field with a complex power spectra has also been reported for the flow of wormlike micelle solutions past a microfluidic cylinder \cite{Haward2019}. In Fig.~\ref{plot}(b), the amplitude of the Beam 1 tip oscillations and the mean beam deflection ($\bar{x}$) are plotted against increasing Weissenberg number. Even though the frequency plateaus beyond $Wi > 50$, the amplitude of oscillations and the mean beam deflection were observed to increase monotonically with the Weissenberg number reaching a maximum amplitude $A = 5 \; \mu$m at the highest Weissenberg number tested, $Wi = 70$.

A second set of experiments was conducted where the flexible Beam 1 was replaced with a more flexible beam, Beam 2 (see details in Table \ref{table}) which had the same beam width and elastic modulus but a longer beam length while maintaining a similar channel blockage ratio. By increasing the beam length while keeping other parameters constant, Beam 2 is an order of magnitude more flexible than Beam 1 resulting in a significantly larger mean deflection under the same flow conditions as seen in the inset of Fig.~\ref{plot}(b). The frequency and amplitude of Beam 2 oscillations are plotted over a range of Weissenberg numbers in Fig.~\ref{plot}(a) and (b). Similar to the case of Beam 1, a re-circulation zone was observed to grow upstream of Beam 2 at low flow velocities while the beam maintained a constant static deflection. The transition from a stable vortex to an unstable time-dependent vortex shedding was observed to occur at a similar Weissenberg number to that for Beam 1, i.e. $Wi = 4$. Accompanying the vortex shedding was the onset of periodic oscillations of Beam 2. The increased flexibility of Beam 2 was observed to have a significant impact on the beam oscillations as the agreement in the critical Weissenberg numbers between the two beams was only possible if the experimentally measured blockage ratio was used to calculate the effective shear rate between the tip of the beam and the bottom wall of the channel. Similar to Beam 1, the frequency of Beam 2 oscillations was observed to increase monotonically before reaching a plateau. For Beam 2, the plateau was observed at a much lower Weissenberg number of $Wi = 12$ (seen in Fig.~\ref{plot}(a) inset). Similar to Beam 1, this plateau in the frequency of oscillations was also found to represent a transition from a coherent vortex shedding to 3D vortex shedding. 

The mechanism of the instability driving the beam oscillations differed slightly in this case of Beam 2 due to its increased flexibility. Unlike the vortex shedding pattern observed for Beam 1, the vortex upstream of Beam 2  was not observed to shed around the tip of the beam en masse. Two streakline images of the flow field upstream of Beam 2 in Fig.~\ref{beam2} show the complex flow conditions occurring during the beam oscillations ($Wi = 8$). The two instances occur at a time interval of $\Delta t = 2.5$ ms. In Fig.~\ref{beam2}(a), a large re-circulating vortex can be observed upstream of Beam 2. Due to the increased beam length, the large vortex is observed to split in two with separate vortices appearing at the tip and the base. The two separate vortices were significantly smaller than the single vortex with some of the fluid shedding around the tip and some flowing through the $4 \; \mu$m gap around the top and bottom of the beam. This alternate vortex shedding pattern was observed during the Beam 2 oscillations upto a Weissenberg number of $Wi = 12$. The appearance of a smaller vortex near the tip of Beam 2 is similar to the lip vortices observed in studies of axisymmetric contraction-expansion flows with rounded corners where rounding of the corner led to a reduction in the contraction ratio and extensional stresses developed in the contraction \cite{boger1994experimental, rothstein2001axisymmetric}.

The amplitude of oscillations of Beam 2 was found to be significantly larger than that of Beam 1, reaching a maximum amplitude of $A= 44 \; \mu$m at $Wi = 8$, but then decaying with increasing Weissenberg number. Due to the increased flexibility of Beam 2, the beam underwent a significant beam deflection with increasing flow velocity. The progression of Beam 2 oscillations with increasing Weissenberg number is presented in Fig.~\ref{longbeam}. The vortex previously observed near the tip of Beam 2 is completely swept off of the beam at these high flow velocities as the beam curvature is unable to support the growth of vortices upstream of the beam and instead, the shear flow along the length of the beam sweeps them off along the beam. The large beam deflection and beam curvature observed in Figs.~\ref{beam2} and \ref{longbeam} are analogous to the curvature of re-entrant corners into planar and axisymmetric contractions \cite{EVANS198995,boger1994experimental,rothstein2001axisymmetric}. As with the rounded re-entrant corners, the beam deflection and curvature resulted in a smoothing out of the streamlines leading to a reduction in the local Weissenberg number \cite{EVANS198995,boger1994experimental,rothstein2001axisymmetric}. This further leads to a reduction of the local extension rate and extensional strain experienced by the polymer solution \cite{ROTHSTEIN199961}. The elastic stress of the fluid passing between the tip of the beam and the opposite channel wall will thus be reduced leading to a decrease in the size of the expected re-circulation zone upstream of the beam and additionally, the amplitude of the beam oscillations. In Fig.~\ref{longbeam}, the displacement of the beam's tip can be seen as a broadening of the beam cross-section in the long time exposure images. Several traces of the flow path of the light-reflective particles are visible in these images. Traces of  particles can be observed approaching the upstream face of the flexible beam, and then moving alongside the length of the flexible beam to slip off the beam edge into the flow. There are also traces of particles visible that move over and under the beam through the small gap between the beam and top and bottom walls. 

The differences in the oscillation amplitudes between the two beams arise from the increased flexibility and the resulting larger mean beam deflection of Beam 2. For a cantilevered beam under a uniformly distributed load, the maximum tip deflection is given by $\delta_{max} = q L^{4} / 8EI$, where $q = F / L$ is a uniformly distributed load \cite{timoshenko1953history}, that stems from the viscous and pressure forces of the viscous flow \cite{Capello, Duprat2014}. The exact pre-factors depend on the specific flow geometry and have been evaluated for the cantilevered beam and the confined geometry \cite{wexler2013bending}. For constant channel and fiber geometry, blockage ratio, elastic and viscous properties, this equation can be simplified to, $\delta_{max} \sim L^{3}$. The Beam 2 deflection will be theoretically scaled by $(l_{2} / l_{1})^{3} = 21 $, where $l_{1}$ and $l_{2}$ are the beam lengths of Beam 1 and Beam 2 respectively. This simplification agrees closely with the mean beam deflection during Beam 2 oscillations observed in the experiments. 

\section{Conclusions}
We report the results of our viscoelastic-fluid structure interaction study of a microscale cantilevered beam subjected to the flow of a polymer solution. The interaction of the elastic flow instabilities with the cantilevered beam was studied on beams with varying flexibility. The flexibility of the beams was modified by increasing the beam length while maintaining the same channel blockage ratio. The critical Weissenberg number at the onset of the spatio-temporal variation of the re-circulation zone upstream of the beam was found to decrease with increasing beam flexibility. The resulting oscillations of the flexible cantilevered beams, triggered by the shedding of the unstable vortex, were observed to display two distinct regimes. The first regime in which the amplitude and frequency of beam oscillations increased with the Weissenberg number was characterized by the shedding of either a single vortex for a less flexible beam or the splitting and then shedding of the upstream vortex as the beam flexibility increased. A second regime was observed where the frequency of oscillations plateaued with increasing Weissenberg number. The onset of this regime occurred much earlier for the more flexible beam case. This second regime of beam oscillations was characterized by 3D chaotic-like instabilities and the absence of a clear upstream vortex. It can be inferred that the evolution of the upstream recirculating zone was coupled with the flow-induced deformation and flexibility of the beam. These experiments provide the first evidence of viscoelastic instabilities triggering motion in flexible cantilevered structures in a confined flow. We have shown that the mechanism of vortex shedding across a flexible structure is heavily influenced by the structural properties such as beam length and flexibility. The critical onset, frequency and amplitude of structural oscillations are a result of the strong coupling between the elastic flow instability and the intrinsic structural flexibility. These conclusions illustrate the complex nature of VFSI and the future possibilities of tuning of the microfluidic flow and/or geometric parameters.

\section*{Conflicts of interest}
There are no conflicts to declare.
\section*{Acknowledgements}
This work was funded by the National Science Foundation under grant CBET-1705251. AL and AAD acknowledge funding from the ERC Consolidator Grant PaDyFlow under Grant Agreement no. 682367 and support from the Institut Pierre-Gilles de Gennes (\'{E}quipement d'Excellence, "Investissements d'avenir", program ANR-10-EQPX-34). The authors would like to thank Jean Cappello and Lucie Duclou\'{e} for their help with the fabrication process of the flexible beams. 

\balance
\bibliographystyle{rsc} 
\bibliography{Draft2} 
\end{document}